\documentclass{elsart}
\usepackage{amssymb}
\usepackage{amsmath}
\usepackage{epsfig}

\begin{document}

\begin{frontmatter}

\title{One abstract method for the study of network transportation and its
application in the traffic problem in the process of city
expansion}

\author{Y.F. Chang\corauthref{email}},
\corauth[email]{Corresponding author. \\
\textit{E-mail address}: yunfeng.chang@gmail.com\\
\textit{Telephone number}: 86-27-67867946}
\author{Xu Cai}

\address{Institute of Particle Physics, Hua-Zhong (Central China) Normal University, Wuhan
430079, China}

\begin{abstract}

One abstract method for the study of network transportation is
proposed in this paper. By interpolating the properties of the
edges that constitute network into the two leading parameters of
the nodes, this method can abstract the configuration of real-life
transportation networks without losing the influence of the
distance and the capability of the physical facilities, which are
essential parts of the transportation networks. With this method
we can study network transportation from the local point of view
and describe network congestion in detail, especially the
dispersing of congestion. After applying this method, as an
example, to the traffic problem in the process of city expansion,
simulation results show that: the intuitions that with the same
rate of particle generation and the same capacity and power, the
smaller the network is the harder for the network to become
congested; and the network will never be congested if the power of
each node equals or is bigger than its capacity are incorrect. And
the distance from one crossroad to another and the width of the
roads can be designed properly, when designing a city, so as to
enlarge the tolerance size of the city in the process of city
expansion. This method may provide more appropriate guide for
real-life traffic design of network transportation.

\bigskip
\noindent{\it PACS}: 89.75.Hc, 89.20.Hh, 89.75.Fb, 05.10.-a
\end{abstract}

\begin{keyword}
 Complex network; Transportation; Congestion; Capability; Power;
\end{keyword}
\end{frontmatter}

\section{Introduction}

The research on the efficiency of transportation on complex
networks is significantly important in different aspects of
natural science and more generally in its practical application.
Many models of traffic flow on complex networks
\cite{s1,s2,s3,s4,s5,s6,s7,s8,s9,s10,s11,s12} can be used to gain
intuition about dynamics on complex networks and to explain the
characteristics of traffic dynamics on networks so as to determine
the leading parameters of network transportation. Currently, the
study of network transportation is mainly focused on network
congestion.

There are two ingredients, the local or global topological
properties of the network and the microscopic dynamical process
involved in the network transportation processes, which are
believed to affect transportation processes \cite{s13,s14}. In
review of the high cost of changing the underlying structure,
researchers prefer to find out optimal strategies for traffic
routing. Nevertheless, researchers prefer to study network
transportation and congestion from the global point of view
without considering the real-life characteristic details of
network transportation.

For simplicity, we will take all kinds of information packets or
cars moving on networks as particles in this paper. In previous
studies \cite{s10,s11,s12,s13,s14}, network congestion was set to
be equal to particle accumulation on networks. An order parameter:
\begin{equation}
\eta(R)=\lim\limits_{t\rightarrow\infty}\langle\Delta
L\rangle/R\Delta t
\end{equation}
or
\begin{equation}
\xi=\lim\limits_{t\rightarrow\infty}W(t)/\lambda Nt
\end{equation}
was used as
the criterion parameter to judge whether the network is congested
or not. There are several shortages with these previous
considerations:

(i) All nodes are given an unlimited capacity \cite{s15} which is
unrealistic.

(ii) Simulations are required a runtime long enough which may
beyond the CPU capability or user-confined simulation steps since
the congestion may need a long waiting-time to occur.

(iii) The congestion of nodes could not be reflected while the
congestion in real-life is node congestion generally.

(iv) The dispersing of congestion in real-life network
transportation systems could not be reflected as well.

(v)  The real configuration of real-life transport network, such
as the influence of the distance and the capability of the
physical facilities, also could not be reflected.

Consequently, previous consideration could not reflect the real
characteristics of network transportation very well, thus it is
difficult to apply previous consideration to solve real
transportation problems.

One abstract method for the study of network transportation from
the local point of view has been proposed in this paper. By
interpolating the real-life configuration properties of the edges
of transportation networks into the two leading parameters of the
nodes: capacity and power, we can abstract transportation network
more reasonably and reflect real-life network transportation in
detail more realistically, especially the network congestion.
After applying this method, as an example, to the traffic problem
in the process of city expansion, simulation results show that:
the intuitions that with the same rate of particle generation and
the same capacity and power, the smaller the network is the harder
for the network to become congested; and the network will never be
congested if the power of each node equals or is bigger than its
capacity \cite{s16} are incorrect. And the distance from one
crossroad to another and the width of the roads can be designed
properly, when designing a city, so as to enlarge the tolerance
size of the city in the process of city expansion. This method may
provide more appropriate guide for real-life traffic design of
network transportation. This method can also be used to study
different kinds of network transportation without losing the
influence of the distance and capability of the physical facility
via which the transportation come into being.

In the next section we will propose our method. In section 3, we
will give out the results of its application in the process of
city expansion and finally in section 4, discussion.

\section{The Method}

\begin{figure}[th]
\centering
\includegraphics[width=6cm]{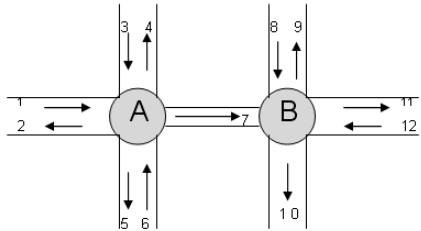} \vspace*{0pt}
\caption{The scheme of crossroads.}
\end{figure}

To interpolate the configuration of real-life transportation
network and make the study of transportation on complex network
more appropriate for real-life transportation, we abstract
networks by including the properties of the edges connected to a
node into the two leading parameters of it: capacity $(C)$ and
power $(P)$. This method abstracts real-life transportation
network to topological structure without losing the real
configuration of transportation network by regarding the
transportation length and the physical facility's power of
delivering/handling particles. The method is implemented in the
following way. Every node has two characteristics: capacity $(C)$
and power $(P)$: the capacity of a node is defined as the maximal
number of the incoming particles that can possibly appear in its
incoming edges per unit time, while the power is the maximal
outgoing particles that can be transported from it via its
outgoing edges per unit time. The unit time could be decided
according to real criterion demanding. In order to describe this
abstract method exactly, we will firstly explain these two
characteristics through an example of city traffic before giving
out the generalized meaning of them.

Take two crossroads depicted in Fig.1 for example, the capacity of
a node is the maximal number of the incoming cars which can
possibly appear in the roads connected to it in one periodical
time of traffic lights. Then every node has its power of
delivering/handling particles which is the maximal number of the
cars that could leave from the roads connected to it in one
periodical time of traffic lights. Here the period of traffic
lights is the unit time. As for the nodes described in Fig.1, for
$A$, the incoming flows $1, 3, 6$ contribute to its capacity while
the out flows $2, 4, 5, 7$ contribute to its power; for $B$, the
incoming flows $7, 8, 12$ contribute to its capacity while the out
flows $9, 10, 11$ contribute to its power. One more thing, the
capacity has relationship with the length and the width of the
roads while the power has relationship with the width of the
roads, thus
\begin{equation}
\begin{array}{clcc}
C(A)=C(1)+C(3)+C(6)\\
P(A)=P(2)+P(4)+P(5)+P(7),
\end{array}
\end{equation}
\begin{equation}
\begin{array}{clcc}
C(B)=C(7)+C(8)+C(12)\\
P(B)=P(9)+P(10)+P(11)
\end{array}
\end{equation}
All of the roads connected to crossroads have their chances to
transport cars in a periodical time of traffic lights. The
smallest power of a node in one periodical time of traffic light
is its outgoing degree, and this smallest power occurs when all
the connected outgoing roads are narrowest with only one driveway.
With the widening of the roads the delivering/handling power of
nodes will increase. What's more, one can easily draw a conclusion
that it is meaningless to have $P>C$ since it is meaningless to
try to deliver/handle more particles than the maximal number of
particles that the node can hold.

In previous studies \cite{s10,s11,s12,s13,s14}, network congestion
was defined to be equal to the accumulation of particles on the
network. But in real-life transportation problems, congestion only
occurs when some particles on the network have nowhere to go. Take
city traffic for example, one rode became congested if there is no
road for the cars in it to go to, that means the congestion of
nodes is caused by the capacity saturation of other nodes but not
that of itself. Here we use a different parameter, congestion time
$T_{c}$, as the criterion parameter of the transportation
capability of networks. The congestion time $T_{c}$ is defined as
the time when the network run into a congestion state that could
not be dispersed anymore. Whereafter, all the nodes in the network
will be congested gradually. With the same rate of particle
generation, $T_{c}$ can reflect the transportation capability of
networks: small $T_{c}$ means the network will be congested
quickly while infinite $T_{c}$ means the network will never be
congested.

This abstract method can be generalized to all kinds of social,
biological, and electronic transportation networks, such as
traffic, internet, WWW, etc., since the efficiency of network
transportation relies on the quality of its physical facilities,
which can affect the maximal number of particles it can hold and
deliver/handle. After interpolating the properties of the edges
that constitute networks into the two leading parameters of nodes
$C$ and $P$, what we need when studying network transportation is
its directed topological structure which determines the incoming
and outgoing degree distribution of the network, and adds in the
network with the intrinsic property of the nodes. And when the
incoming and outgoing by edges could not be distinct exactly, the
capacity was added with a renormalization condition.

\section{Results and Analysis}

Below are the simulation results when we apply this abstract
method to the traffic problem in the process of city expansion.

\begin{figure}[th]
\centering
\includegraphics[width=5cm]{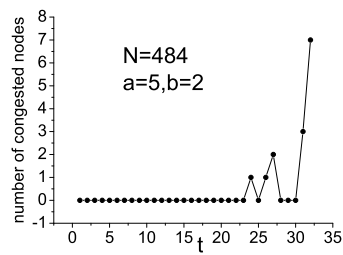} \vspace*{0pt}
\caption{The number of congested nodes versus time.}
\end{figure}

When simulation, we adopt a linear relationship between the
capacity/power and the incoming degree/outgoing degree: $C=a\times
k_{in}, P=b\times k_{out}$, where $k_{in}$ is the incoming degree
and $k_{out}$ is the outgoing degree, $a$ and $b$ are two tunable
parameters which can reflect the length and the width of the
roads. And we base our simulation on a most simple two-dimensional
lattice by regarding that two-dimensional lattice is close to the
real structure of city traffic network and the roads are mostly
two-way roads, namely $k_{in}=k_{out}$. Particles can incoming or
outgoing via all of the roads along the shortest path, as well as
steer clear of the congested nodes. The number of particles
generated at every time step $n=N/2$, $N$ is the number of nodes
in the network.

In Fig.2, the dispersing of congested nodes can be observed. The
network can only revert to free flow when the number of congested
nodes is very small because of the big rate of particle generation
at every time step, and the congestion time $T_{c}=31$.

\begin{figure}[th]
\centering
\includegraphics[width=4cm]{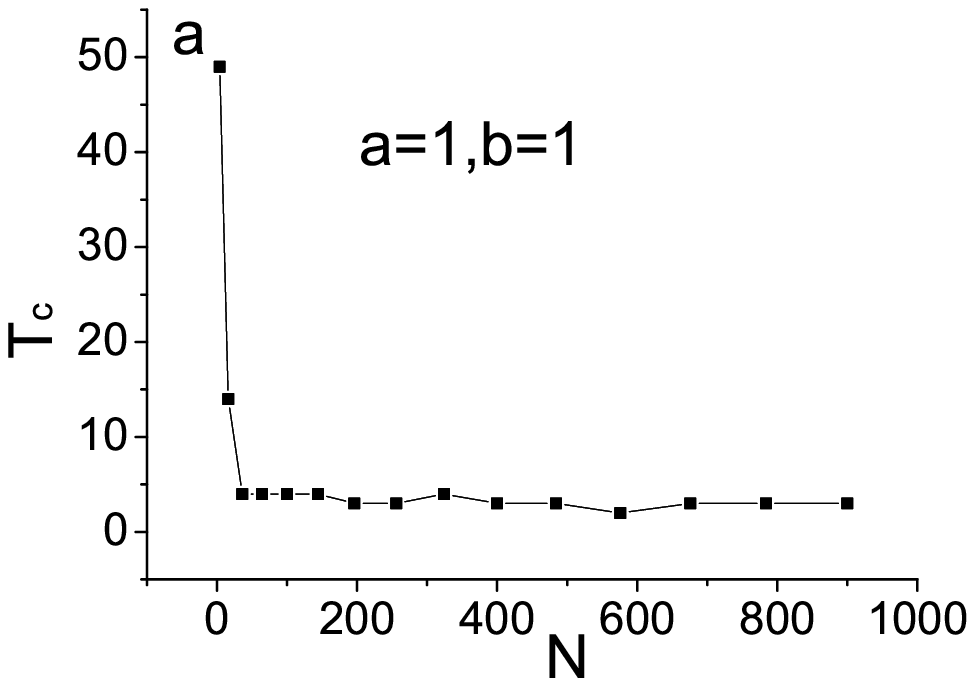} \vspace*{0pt}
\includegraphics[width=4cm]{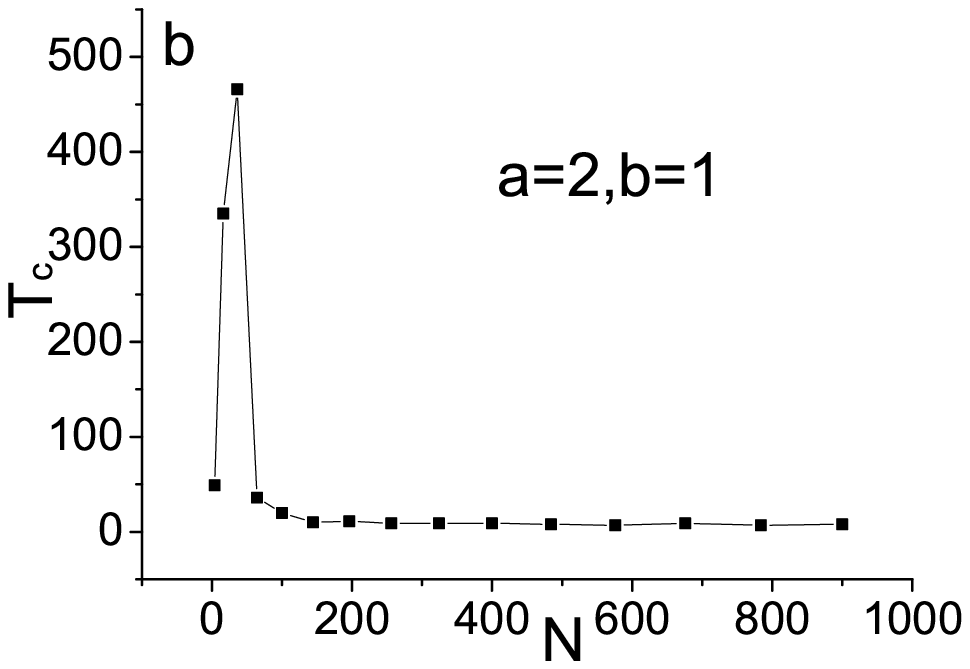} \vspace*{0pt}
\includegraphics[width=4cm]{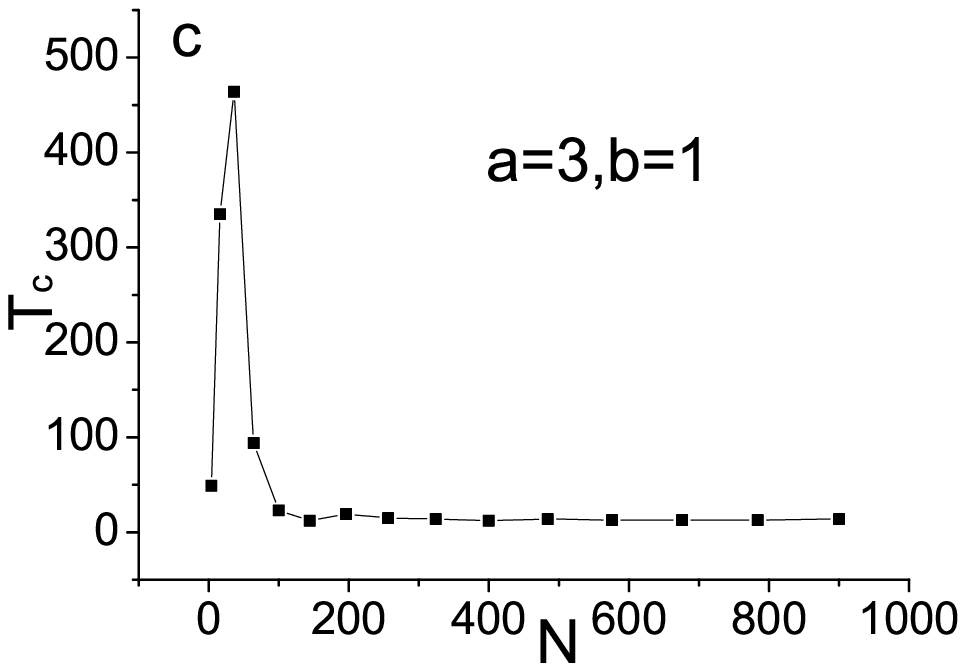} \vspace*{0pt}
\includegraphics[width=4cm]{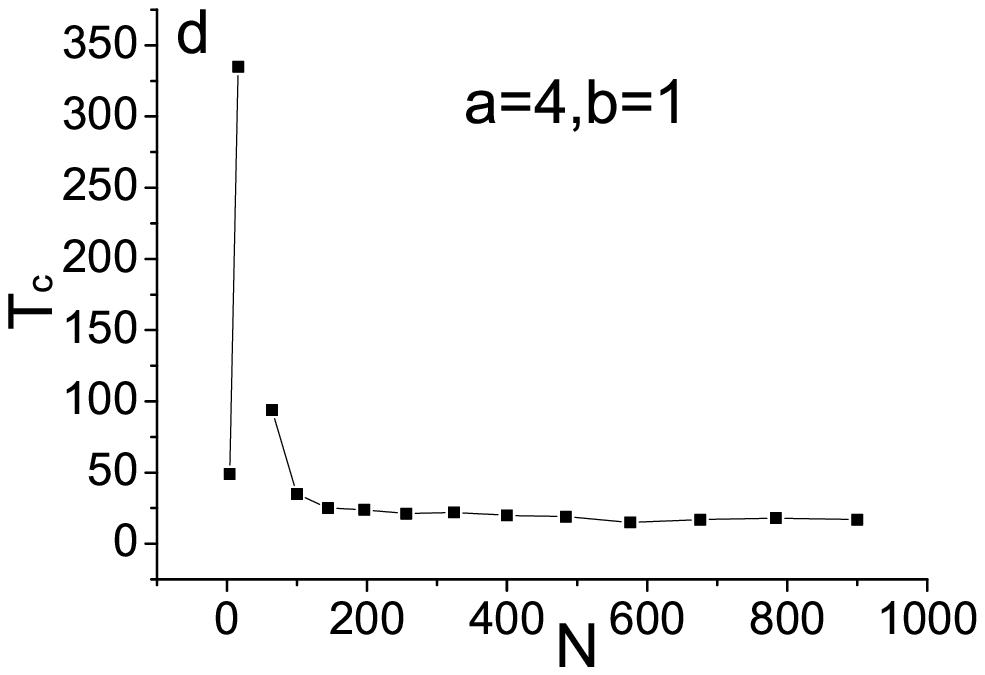} \vspace*{0pt}
\includegraphics[width=4cm]{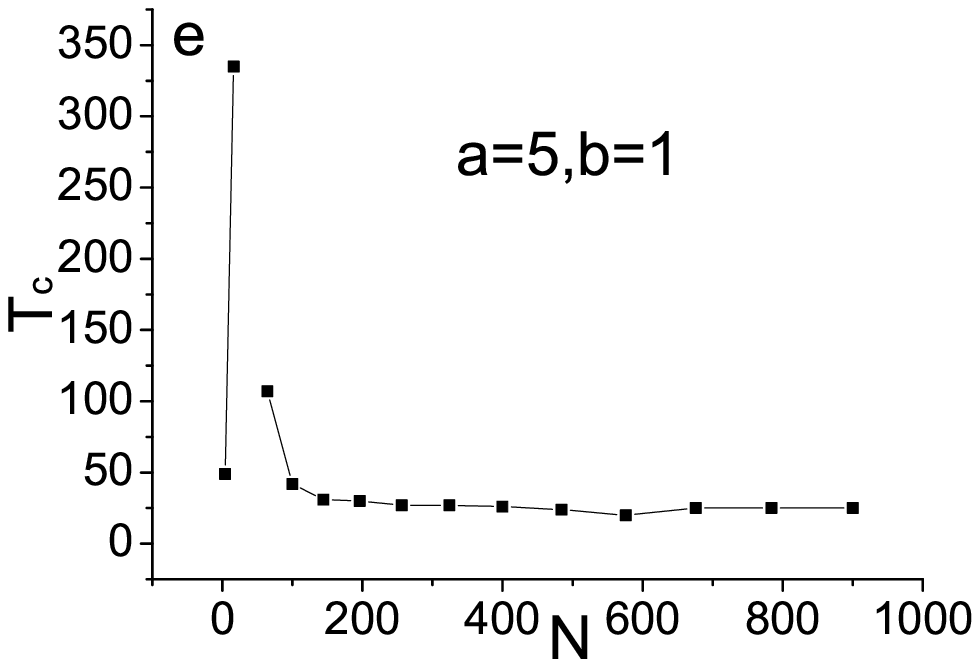} \vspace*{0pt}
\includegraphics[width=4cm]{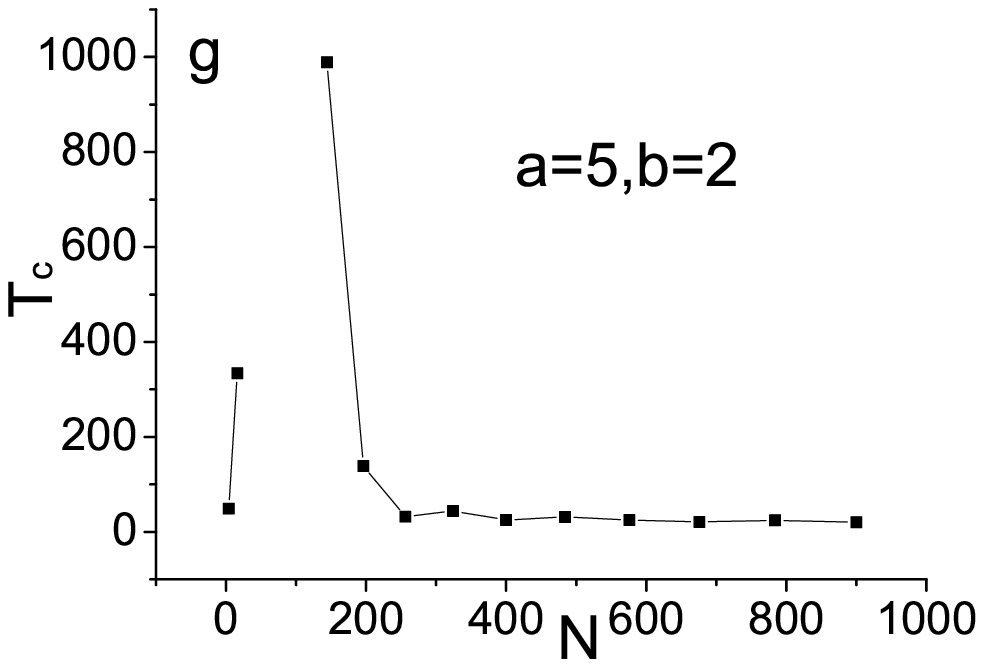} \vspace*{0pt}
\includegraphics[width=4cm]{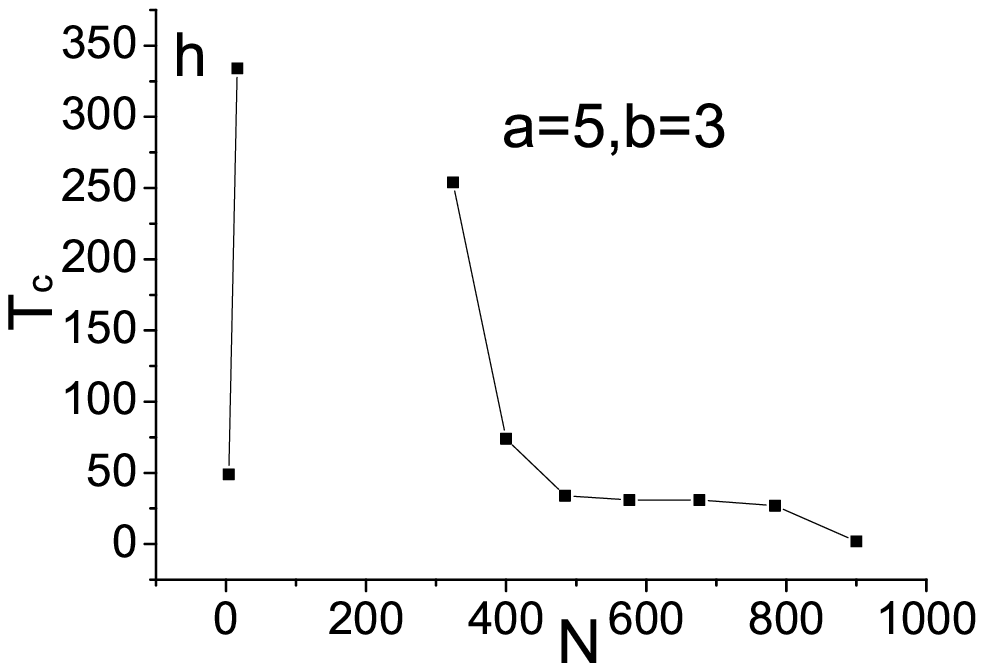} \vspace*{0pt}
\includegraphics[width=4cm]{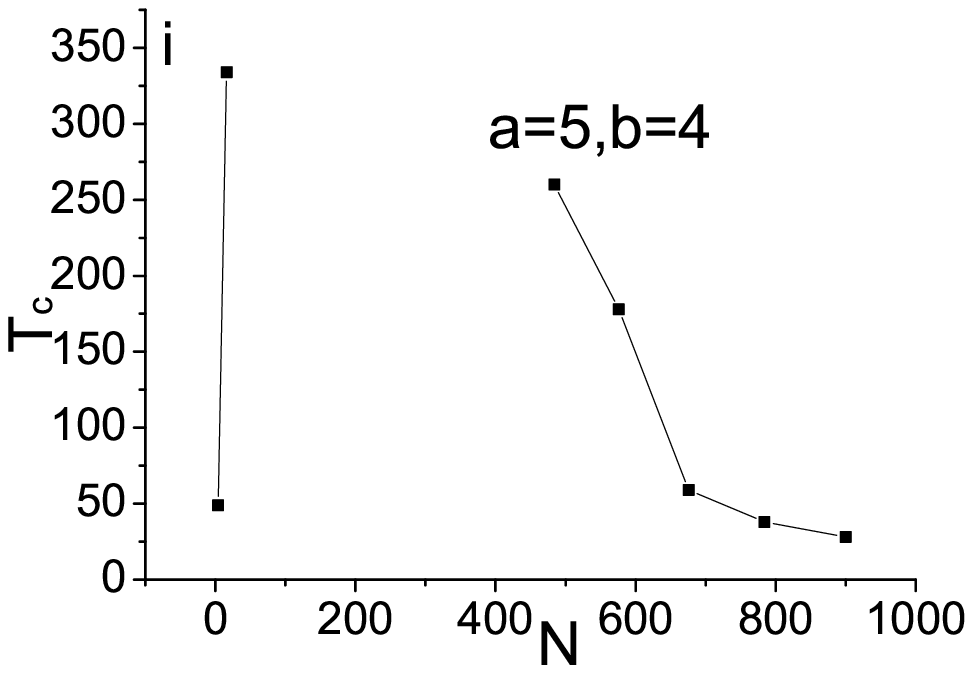} \vspace*{0pt}
\includegraphics[width=4cm]{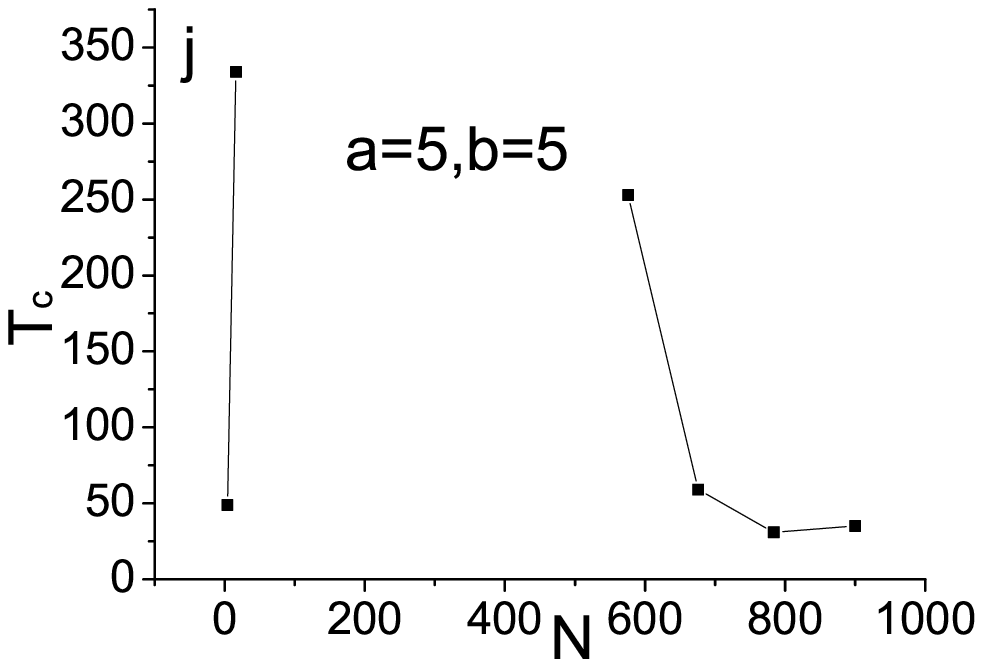} \vspace*{0pt}

\caption{The congestion time versus the size of the network.}
\end{figure}

The size of the networks in Fig.3 goes from $2\times2$ to
$30\times30$. As explained above, the smallest value of $b=1$, and
the gap in Fig.3(d-j) means the congestion time exceeds our
simulation time steps. Fig.3 shows that:

(i) With the same rate of particle generation and the same
capacity and power, the intuition that the smaller the network is
the harder for the network to become congested is proved to be
incorrect. In fact this situation occurs only when the
transportation network has the smallest capacity and power.
Otherwise, with the increase of capacity or power, certain values
of $a$ and $b$ correspond to networks with certain size that has
the most outstanding transportation capability but not the smaller
or bigger ones.

(ii) Another intuition that the network will never be congested if
the power of each node equals or is bigger than its capacity
\cite{s16}, namely $P\geq C$, is also proved to be incorrect
(Fig.3 (j)). This account for the fact that congestion is caused
by the situation that there is nowhere for some particles to go
but not the saturation of capacity, as explained above.

(iii) The width of the gap in Fig.3 (e-j) is increasing, that
means the values of $a$ and $b$ can be set properly by considering
the distance from one crossroad to another and the width of the
roads connected to them, when designing city traffic, so as to
enlarge the tolerance size of the city in the process of city
expansion, consequently avoid frequent reconstruction or
over-estimated building of city roads which is a waste of natural
resource and human power.

What's more, the simulation results show that the variation of
power is more efficient than the variation of capacity to change
the transportation capability.

\section{Discussion}

In summary, one abstract method for the study of network
transportation from the local point of view has been proposed in
this paper. By interpolating the real-life configuration
properties of the edges of transportation networks into the two
leading parameters of the nodes: capacity and power, we can
abstract transportation network more reasonably and reflect
real-life network transportation in detail more realistically,
especially the network congestion. You can increase the rate of
particle generation per time step, thus avoid the missing of
congestion because of the limited simulation time or CPU time,
without losing the intrinsic property of the network
transportation. After applying this method to the traffic problem
in the process of city expansion, we get some significant results
that have not been reflected in previous studies, and may provide
more instructive guide for the traffic design of network
transportation. This method can also be used to study different
kinds of network transportation without losing the influence of
the distance and capability of the physical facility via which the
transportation come into being.

What's more, we will consider nonlinear or more complex
relationship between the capacity/power and the directed degree,
as well as changeable rate of particle generation in our future
work.

\section*{Acknowledgements}

We thank Liang Sun and Qiong Lei for discussion and helpful
comments. This work is supported by the National Natural Science
Foundation of China under grant no. 70571027, 70401020, 10647125
and 10635020, and by the Ministry of Education of China under
grant no. 306022.


\begin{thebibliography}{s99}

\bibitem{s1} R. Guimer$\acute{a}$, A. D$\acute{i}$az-Guilera, F.
Vega-Radondo, A. Cabrales, and A. Arenas, Phys. Rev. Lett. {\bf
89}, 248701 (2002).
\bibitem{s2} B. Tadic, S. Thurner, and G. J. Rodgers, Phys. Rev. E {\bf 69}, 036102 (2004)
\bibitem{s3} P. Echenique, J. Gomez-Garde$\tilde{n}$es, and Y. Moreno, Phys. Rev. E {\bf 70}, 056105 (2004)
\bibitem{s4} L. Zhao, Y.C. Lai, K. Park, and N. Ye. Phys. Rev. E {\bf 71}, 026125 (2005)
\bibitem{s5} B. K. Singh and N. Gupte, Phys. Rev. E {\bf 71}, 055103(R) (2005)
\bibitem{s6} K. I. Goh, J. D. NOh, B. Kahng, and D. Kim, Phys. Rev. E {\bf 72}, 017102 (2005)
\bibitem{s7} M. C. Santos, G. M. Viswanathan, E. P. Raposo, and M. G. E. da Luz, Phys. Rev. E {\bf 72}, 046143 (2005)
\bibitem{s8} L. Donetti, P. I. Hurtado, and M. A. Mu$\tilde{n}$oz, Phys. Rev. Lett. {\bf 95}, 188701 (2005)
\bibitem{s9} D. J. Ashton, T. C. Jarrett, and N. F. Johnson, Phys. Rev. Lett. {\bf 94}, 058701 (2005)
\bibitem{s10} W. X. Wang, B. H. Wang, C. Y. Yin, Y. B. Xie, and T. Zhou, Phys. Rev. E {\bf 73}, 026111 (2006)
\bibitem{s11} G. Yan, T. Zhou, B. Hu, A. Q. Fu, and B. H. Wang, Phys. Rev. E {\bf 73}, 046108 (2006)
\bibitem{s12} Z. Y. Chen and X. F. Wang, Phys. Rev. E {\bf 73}, 036107 (2006)
\bibitem{s13} Z. Y. Chen, and X. F. Wang, Phys. Rev. E {\bf 73}, 036107 (2006)
\bibitem{s14} W. X. Wang, C. Y. Yin, G. Yan, and B. H. Wang, Phys. Rev. E {\bf 74}, 016101 (2006)
\bibitem{s15} The maximal number of particles a node can hold.
\bibitem{s16} Jordi Duch, and Alex Arenas, Phys. Rev. Lett. {\bf 96}, 218702 (2006)

\end{thebibliography}
\end{document}